# A Cost-Effective Upgrade Path for the Fermilab Accelerator Complex

S. Nagaitsev[1] and V. Lebedev, Fermilab, Batavia, IL 60510, USA

Abstract

The Fermilab Proton Improvement Plan II, or PIP-II, would enable the world's most intense high-energy neutrino beam and would help scientists search for rare particle physics processes. The PIP-II goal is to deliver 1.2 MW of proton beam power from the Fermilab Main Injector, over the energy range 60 – 120 GeV, at the start of operation of the LBNF/DUNE program. PIP-II provides a variety of upgrade paths to higher beam power from the Main Injector, as demanded by the neutrino science program and as recommended by the 2014 P5 report. Delivering more than 2 MW to the LBNF target in the future will require a replacement of the existing Booster. This report outlines a cost-effective Booster replacement option and an upgrade path for the Fermilab Accelerator Complex to attain 2.4-MW beam power on the LBNF target, as well as to retain the capability to provide 8-GeV proton beams to the existing Fermilab Muon Campus via the existing Recycler ring. Its cost-effectiveness is achieved by: (1) using a small-diameter metallic vacuum chamber in the Booster replacement and (2) reusing the existing Recycler ring. Reusing the Recycler ring may be of particular advantage since it is presently employed to deliver 8-GeV beams to the Muon Campus experiments. The present concept also retains such a capability.

---

[1]Also at the University of Chicago, Chicago, IL 60637, USA

# I   Introduction

The present Fermilab proton Booster, operating at 15 Hz, is an early example of a rapidly-cycling synchrotron (RCS). Build in the 1960s, it features a design in which the combined-function dipole magnets serve as vacuum chambers for the beam. Such a design is quite cost-effective, and it does not have the limitations associated with metallic vacuum chambers having eddy currents due to rapidly changing magnetic fields of the RCS during beam acceleration. However, an important drawback of such a design is a high impedance, as seen by the beam, because of magnet laminations. More recent RCS designs (e.g. the J-PARC RCS operating at 25 Hz), employ large and complex ceramic vacuum chambers in order to mitigate the eddy-current effects and to shield the beam from magnet laminations. Such a design, albeit very successful, is quite costly because it requires complex ceramic vacuum chambers, large-bore magnets and large-bore rf cavities. In this report, we propose an RCS concept with a thin-wall metallic vacuum chamber as a compromise between the chamber-less Fermilab Booster design and a large-bore design with ceramic chambers. Due its conductivity, a thin-walled metallic vacuum chamber still has eddy-current effects, but these effects can be easily mitigated as presented below. We propose to use a vacuum chamber made of 316LN stainless steel; however, the Inconel 718 alloy may offer several technical advantages over the 316LN construction although at a much high cost for the material.

The most straightforward upgrade path for the Fermilab accelerator complex would be to extend the 0.8-GeV PIP-II linac to 1.5-2 GeV and to inject at this energy into a new 8-GeV rapid cycling synchrotron (RCS). In this report, we will describe a concept (see Refs [1, 2]) of a 2-8 GeV proton RCS with a primary goal to deliver 2.4 MW beam power from the Fermilab Main Injector to the LBNF neutrino target. For this concept, we will assume a 2-GeV, 2-mA H- linac as an injector, capable of bunch-by-bunch chopping for the longitudinal painting at injection into the RCS. We will also assume a fixed-energy (8 GeV) accumulation ring (the existing Fermilab Recycler ring) to collect several RCS batches during the Main Injector 8-120 GeV ramp. We account that the Recycler and the Main Injector have equal circumferences, thus, allowing for a single-turn bucket-to-bucket transfer. Figure 1 schematically shows the proposed Fermilab site layout.

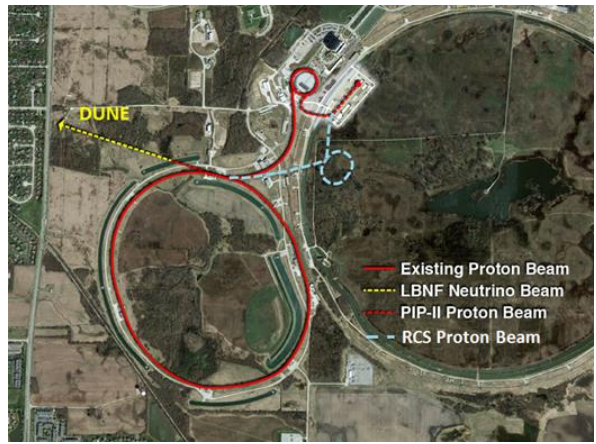

**Figure 1**: The schematic RCS site layout.

## II  RCS design parameters

To support the 2.4-MW operation of the Fermilab Main Injector (MI) for the LBNF/DUNE program, a rapid cycling synchrotron (RCS) must deliver ~1.6·10$^{14}$ protons to the MI on each cycle, which can vary in length from ~0.7 sec (for the 60 GeV extraction energy) to ~1.2 sec (120 GeV).  The RCS has a shorter circumference than the Recycler and therefore several RCS cycles are required to fill it.  Balancing the impacts of beam space charge at injection, instabilities, the magnetic field strength, and the repetition rate, the circumference is chosen to be 1/6 of the MI circumference and the repetition rate is chosen to be 10 Hz. There are 12 RCS cycles a 1.2-sec MI cycle, of which 6 are used for the 2.4-MW neutrino program and 6 remaining cycles are available for the Muon Campus or other 8-GeV programs, operated from the Recycler.  The main parameters are presented in Table I.

**Table I: Main Parameters of RCS**

| Energy, min/max, GeV | 2/8 |
|---|---|
| Repetition rate, Hz | 10 |
| Circumference, m (MI/6) | 553.2 |
| Tunes, $v_x/v_y$ | 18.42 / 18.44 |
| Transition energy (kinetic), GeV | 13.3 |
| Number of particles (extracted) | 2.6 x 10$^{13}$ |
| Beam current at injection, A | 2.2 |
| Transverse 95% normalized emittance at injection, mm mrad | 22 [2] |
| Space charge tune shift, at injection | 0.1 [3] |
| Normalized acceptance at injection, mm mrad | 40 |
| Harmonic number for main RF system, $h$ | 98 |
| RF bucket size at injection, eV-s | 0.38 |
| Injection time for 2-mA linac current, ms | 2.1 |
| Total beam power from linac, kW | 87 [4] |
| Total beam power delivered by RCS, kW | 340 |

The requirements for reliable and efficient operations of the RCS lead to some specific design decisions.  To avoid transition crossing during RCS ramping, the transition energy is chosen to be outside of the machine energy range.  The focusing lattice needs to be relatively insensitive to focusing errors and synchro-betatron resonances. The latter is achieved by having a zero dispersion function in the RCS straight sections, where RF cavities are located.  A FODO lattice with a missing dipole for dispersion suppression is a natural choice, resulting in modest requirements for magnets and vacuum chambers as well as reduced sensitivity to nonlinearities.  However, to avoid the stripping foil overheating by the injected beam, the beta-functions at the foil location need to be increased relative to their nominal values of the FODO structure. Therefore, the optics for seven half-cells near the injection point was modified. This results in an increase of the geometric mean of beta-functions at foil, $\sqrt{\beta_x \beta_y}$, from 5.5 m to 20.5 m with the corresponding decrease of foil heating by more than an order of magnitude. Note, that if such large beta-function values were used in the arcs, it would result in the transition energy being within the RCS energy range regardless of what kind of single cell optics is used (FODO, doublet, triplet, etc.). It would also result in an increase of the beam pipe size and its eddy-current heating by dipoles.

---

[2] A 12% emittance dilution is assumed during acceleration in the RCS and transfer to Recycler, where the design value for the emittance is 25 mm mrad.

[3] This value is estimated for a KV-like transverse distribution and the longitudinal bunching factor of 2.2 that are obtained by the beam painting as presented below in Section 5.

[4] We imply here a 4% loss of the injected beam. See details below.

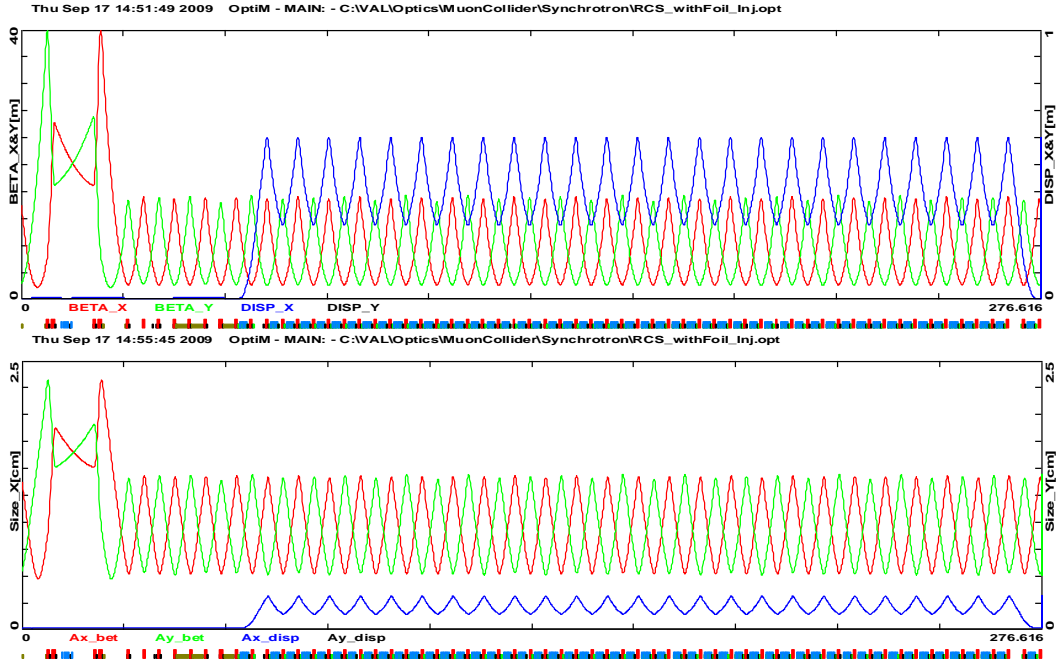

**Figure 2**: The beta-functions and dispersion (top) and the beam envelopes (bottom) for a half of the ring (the injection & extraction straight and downstream arc.) The beam envelopes are shown at injection for $\varepsilon_n$=40 mm mrad, $E_k$ = 2 GeV, $\Delta p/p$ = 5 x 10$^{-3}$.

The RCS is designed as a racetrack (two long straight sections and two 180° arcs) with the same distance between the centers of quads through the entire ring with exception of the injection region. One long straight section is used for the RF cavities and the other one is used for injection, extraction, and beam collimation. The F and D quads have the same focusing strength and are connected serially with the dipoles. Eight quadrupoles, four in the injection and four in the extraction regions, have a larger aperture and length, but they have the same integral strength. The tune and optics corrections are performed via additional corrector coils wound in each quadrupole. Figure 2 presents the ring beta-functions, the dispersion, and the beam envelopes for one half of the ring. The injection region is shown on the left side of the plot, where two quadrupole doublets replace six quadrupoles of the FODO structure. The rest of the ring, including another straight line not shown in the picture, has the regular FODO structure. The betatron phase advance per cell is 102°. Strong focusing results in small beam sizes and small dispersion. That, in its turn, results in a small disperion-related beam size and, consequently, a small difference between horizontal and vertical beam envelopes through the entire ring.

The RCS FODO focusing structure is designed so that for a perfectly periodic lattice it would have 66 cells with two 7-cell long straight sections. Considering that 7 half-cells of the injection straight are used for injection optics and they have 4 quads instead of 6, one obtains that altogether there are 130 quadrupoles and 100 dipoles. To ease the magnet power supply voltage requirements, the dipoles and the quadrupoles of all cells are included into a 10-Hz resonance circuit. Every quadrupole has an associated corrector package. The general corrector package contains a trim dipole coil (horizontal near the F quads and vertical near the D quads) and a sextupole coil at the locations with non-zero dispersion.

There are 98 sextupoles in the RCS ring concept. They are located at the positions 16-64 and 82-130 and are split into two families F (50 sextupoles) and D (48 sextupoles). The sextupole strengths required to correct the natural machine chromaticities, $\xi_x \approx \xi_y \approx -25$, to zero are $\int S_F dL$ = 3.7 kG/cm and $\int S_D dL$ = -6.8 kG/cm at the beam energy of 8 GeV. Considering that the RCS will be operated with chromaticities in the ranges of -10 to -20 we chose the maximum strength to be 4 kG/cm. This still leaves a sufficiently large margin between the operational and maximum strengths of sextupoles.

The relatively low beam current of the injection linac, 2 mA, results in long injection time, 2.1 ms or ~1000 turns. To minimize the required beam energy correction, the injection begins 1.05 ms before the magnetic field reaches its minimum (assuming the 10-Hz RCS cycle). The corresponding variation of the bending field of ±0.11% is compensated by dipole correctors.

## III  RCS vacuum chamber

The RCS vacuum chamber needs to satisfy a set of competing requirements. To minimize the effects of eddy currents in a conductive beam pipe, a thin-wall small-radius vacuum chamber would be optimal. To minimize the transverse impedance, a high-conductivity wall with a large-radius pipe would be optimal. In summary, the competing effects are:
- shielding and distortion of the dipole bend field by eddy currents, excited in the vacuum chamber;
- vacuum chamber mechanical integrity under the atmospheric pressure;
- vacuum chamber heating by eddy currents;
- transverse beam impedance due to the wall resistivity;
- geometric ring acceptance.

The compromise resulted in a round stainless-steel 316LN vacuum chamber with the external radius of 22 mm and the wall thickness of 0.7 mm. Note that in the RCS arcs (Figure 2), the maximum beam radius is ~15 mm.

The complex amplitude of the magnetic field, generated by eddy currents, which are excited in a cylindrical vacuum chamber of radius $a_w$ with the wall thickness $d_w$ and the wall conductivity $\sigma_w$ by the AC component of dipole magnetic field, alternating with frequency $\omega_{ramp}$ is:

$$\Delta B_y(0, y) = iB_{AC}\left(1 + \frac{\pi^2}{12} + \frac{\pi^4}{240}\frac{y^2}{a_w^2} + ...\right)\frac{a_w d_w}{\delta_w^2}, \quad \delta_w = \frac{c}{\sqrt{2\pi\sigma_w\omega_{ramp}}} >> a_w, \qquad (1)$$

where $\delta_w \approx$ 13.8 cm is the skin depth in the stainless steel 316LN at 10 Hz. One can see that this field correction is phase-shifted by 90 deg. relative to the AC component of the dipole magnetic field. The first addend in the parenthesis is related to the eddy currents, excited in the cylindrical chamber wall by the uniform dipole field; and the other two addends are related to the multiple image currents of this wall current in the poles of a dipole, assumed to be separated by $\sim a_w$. For the magnetic field changing as $B(t) = B_{DC} - B_{AC}\cos(\omega_{ramp}t)$, the last addend corresponds to the sextupole field with the sextupole strength equal to:

$$S(t) = B_{AC} \frac{\pi^4}{120} \frac{d_w}{a_w \delta_w^2} \sin(\omega_{ramp}t) . \tag{2}$$

The relative value of the dipole field correction (the sum of the first two addends in Eq. (1)) is equal to zero at the cycle start and end. It achieves its maximum of $|\Delta B/B|=8.5 \cdot 10^{-4}$ at 16 ms within the acceleration cycle. Similarly, in the case of a changing quadrupole field in a quadrupole magnet there is a quadrupole field correction with a relative value approximately a half of the dipole correction. Consequently, keeping constant tunes during the acceleration cycle requires a quadrupole current correction $\Delta I/I \approx 4.3 \cdot 10^{-4}$. Like the dipole field correction, the sextupole field correction is zero at the cycle beginning and end. At 16 ms it results in a maximum contribution to the machine chromaticity: $\Delta \xi_x \approx 1.03$ and $\Delta \xi_y \approx -0.85$. These values are a small fraction of the natural machine chromaticity and can be easily compensated by the machine sextupoles. Note that the sign of the sextupole correction is "focusing' in the acceleration part of the cycle and "defocusing" in the deceleration part. Tracking studies show that if only the machine sextupoles and the eddy current sextupoles are considered, the dynamic aperture exceeds the machine aperture by about a factor of 4.

There must be a sufficiently large safety margin for the mechanical stresses in the vacuum chamber to ensure reliable operations for a long lifetime of the machine. For a perfectly cylindrical vacuum chamber, the stress in the material due to the atmospheric pressure, $P_{atm}$, is determined as

$$\sigma_{cmpr} = P_{atm} \frac{a_w}{d_w} , \tag{3}$$

and is equal to 3.1 N/mm². If the chamber is slightly elliptic there is an additional stress, equal to

$$\sigma_{bend} = \frac{9}{4} P_{atm} \frac{\Delta a_w}{a_w} \left(\frac{a_w}{d_w}\right)^2 , \tag{4}$$

where $\Delta a_w$ determines the ellipse semi-axes to be equal to $a_w \pm \Delta a_w$. Assuming a comparatively conservative ellipticity of the chamber $\Delta a_w / a_w = 0.02$, corresponding to $a - b = 0.88$ mm, we obtain $\sigma_{bend} = 8.9$ N/mm². The sum of these two stresses is equal to 12 N/mm². It is ~20 times smaller than the maximum acceptable stress (the yield stress) for the 316LN stainless steel of ~200 N/mm². A further reduction of the vacuum chamber thickness is still possible but a thickness below 0.5 mm would jeopardize the long-term stability and the integrity of the chamber.

The eddy currents produce the vacuum chamber heating load because of ohmic losses. The thermal power per unit length is

$$\frac{dP}{dz} = \frac{\pi \sigma_R d_w a_w^3 \omega_{ramp}^2}{c^2} B_{AC}^2 . \tag{5}$$

For the nominal RCS parameters this heating load is equal to ~10 W/m. This power level does not require water or forced air cooling. A conservative air-cooling estimate for the case of convective cooling is based on the heat transfer coefficient of $10^{-3}$ W/cm$^2$/K. If one neglects the thermal conductivity of the chamber it results in a temperature increase of 15 C (compared to ambient) on the both sides of the chamber, where the current is concentrated. Further increase of the ramp frequency or the AC component of magnetic field would require forced air cooling.

As discussed above, the proposed beam vacuum tube diameter for the quadrupoles and the dipoles is 44 mm OD, transitioning to a standard 2-inch OD tube for the rest of the vacuum chamber. The wall thickness is 0.7 mm inside the magnets. It can be larger at other places. Bellows and pumping ports will require RF shielding. In order to minimize the secondary electron emission yield from the chamber wall, the entire vacuum chamber will be coated with TiN. This has been done successfully at the SNS and the PEP II B-Factory. The vacuum chamber will be constructed of the 316LN stainless steel. It will be electropolished and hydrogen degassed prior to coating with TiN. Alternatively, Inconel 718 can be considered as a promising choice of material for the quadrupole and the dipole beam tubes. Compared to the stainless steel, it has a higher electrical resistivity, which reduces the eddy current effects, and a higher strength for structural stability. Disadvantages include the high cost and the challenge of coating with a thin and highly uniform copper layer to reduce impedances, which is not be required for the 316LN stainless-steel option.

Potentially, the non-linear eddy current fields excited in a chamber could be passively compensated with correction windings on the vacuum chamber wall like it was implemented at the Brookhaven AGS Booster [3]. The scheme successfully corrects the non-linear field for arbitrary ramp rates and allows for reasonably thick-walled, low impedance and possibly large aperture vacuum chambers. However, taking into account relatively small non-linearity introduced by eddy currents in the considered above vacuum chamber and additional thermal effects associated with the coils we consider such addition being excessive.

## IV  RCS impedances and beam stability

At low frequencies (f ≤ 0.5 GHz), the transverse impedance is dominated by a direct beam interaction the with vacuum chamber and its the wall resistivity. Figure 3 presents the real and imaginary parts of the transverse impedance for the above considered thin wall round vacuum chamber at the injection energy. The impedance accounts for both the wall resistivity and the direct electro-magnetic beam interaction with the vacuum chamber. The latter is negligible at the ultra-relativistic energy but becomes significant at a low energy. In particular, it results in that the imaginary part stops to decrease with increasing frequency as can be seen in Figure 3 above 100 kHz. As one can see, the real part of the impedance decreases with frequency as $1/\sqrt{\omega}$. However, when the skin-depth becomes larger than the vaccun chamber thickness, the dependence on frequency becames $1/\omega$. The lowest betaron side-bands belong to this area, where the real part of the impedance per unit length can be approximated by the following equation:

$$Z_\perp(\omega) = Z_0 \frac{c^2}{4\pi^2 \sigma_R \omega a_w^3 d_w} \quad , \quad \delta_w = \frac{c}{\sqrt{2\pi\sigma_w \omega}}, \quad \sqrt{a_w d_w} \geq \delta_w \geq d_w \quad . \tag{6}$$

Here $Z_0 \approx 377$ Ω is the impedance of vacuum. Comparing Eqs. (5) and (6) one can see that the transverse impedance and the vacuum chamber heat load are closely related:

$$Z_\perp(\omega)\frac{dP}{dz} = \frac{Z_0}{4\pi} \frac{\omega_{ramp}^2}{\omega} B_{AC}^2 \quad . \tag{7}$$

Their product does not depend on the vacuum chamber radius, thickness and material. Varying the parameters to reduce the heating term results in a corresponding increase of the impedance and vice versa.

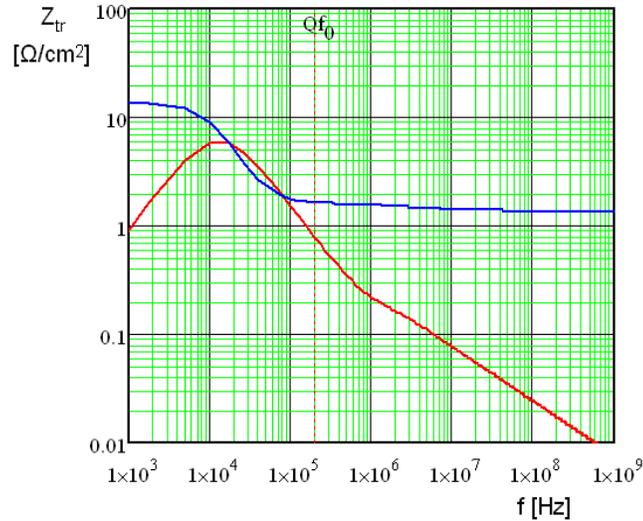

**Figure 3**: The dependence of real (red) and imaginary (blue) parts of transverse impedance per unit length on the frequency for round stainless steel vacuum chamber: radius of 22 mm, wall thickness of 0.7 mm, and the beam energy 2 GeV.

Figure 4 presents the betatron tune shifts due to the beam interaction with the stainless-steel vacuum chamber in the approximation of a zero betatron tune spread. The growth time for the lowest betatron sideband is about 150 turns. A bunch-by-bunch transverse damper can suppress this instability easily. Instabilities at frequencies corresponding to intra-bunch motion ($f > 50$ MHz) are damped by the chromaticity (see below).

As one can see from the above discussion, there are many reasons to keep the vacuum chamber size being sufficiently small. It also shows that a ceramic vacuum chamber is not really beneficial (while significantly more expensive): a reduction of conducting layer thickness, required to shield the ceramic from beam, could reduce the vacuum chamber heating but it results in an increase of the transverse impedance being the same as for a stainless-steel chamber. If the vacuum chamber wall is too thin or even absent as in the Fermilab Booster, the beam interacts with the steel laminations of magnets, resulting in much higher impedances and instability growth rates.

There is a sufficiently large margin between the 95% normalized emittance of 25 mm-mrad and the beam boundary set at 40 mm-mrad. At injection the beam boundary corresponds to the maximum beam size of 14 mm leaving 6 mm for orbit distortions in the both planes. This value is within normal operational orbit distortions of the present Fermilab Booster. The sagitta in the dipole is equal to 1.7 cm. To avoid aperture loss, the vacuum chamber in the dipoles must be bent to follow the beam bending with 33.9-m radius.

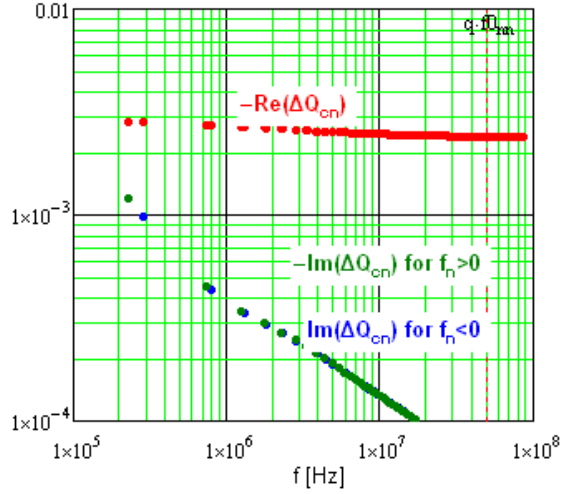

**Figure 4:** Real and imaginary tune shifts for the stainless-steel vacuum chamber; $a_w$=22 mm, $d_w$=0.7 mm, 2 GeV beam energy.

A beam is stable, if for its every coherent mode a sum of Landau damping and the damper-induced rates exceeds the impedance-associated growth rate. Landau damping is extremely sensitive to the ratio of the effective space charge tune shift to the synchrotron tune [4]. For the discussed parameters, this ratio is $q$=2.4 for a 3D Gaussian beam, and it is $q$=1.6 for KV model at injection energy. Since the space charge parameter $q$ is not extremely high, the strong space charge theory [4] can be used only as a rather rough approximation. According to that, the 0$^{th}$ head-tail mode does not have any Landau damping at all, while the first mode has rather fast damping rate $\Lambda_1 \approx (0.2-0.4)Q_s$, and higher modes should not be seen at all.

The coherent growth rate can be estimated by the air-bag model (Ref. [5], Eq. (6.188)]:

$$\mathrm{Im}\,\Omega = \frac{N r_0 c}{2\gamma T_0^2 \omega_b} \sum_{p=-\infty}^{\infty} Z_\perp(\omega') J_l^2(\omega'\hat{z}/c - \chi) \,, \quad \omega' = pM\omega_0 + \mu\omega_0 + \omega_b. \tag{8}$$

For the resistive wall impedance, given by Eq. (6), it results in a growth rate equal to:

$$\mathrm{Im}\,\Omega/\omega_0 = \frac{N r_0 \delta_0 \bar{\beta}_x}{2\pi \gamma a^3} \Gamma(l,\chi,\mu)\,;$$

$$\delta_0 = c/\sqrt{2\pi\sigma\omega_0}\,; \tag{9}$$

$$\Gamma(l,\chi,\mu) \equiv \sum_{p=-\infty}^{\infty} \sqrt{\frac{\omega_0}{|\omega'|}} J_l^2(\omega'\hat{z}/c - \chi)\,\mathrm{sgn}(\omega')$$

The most unstable coupled-bunch mode for the betatron tune $\nu_b$=18.44 is the mode $\mu$ = -19. The mode-factors $\Gamma$ for this coupled-bunch number are presented in Figure 5. Without a damper, the 0$^{th}$ head-tail mode can be only stabilized for the head-tail phase $\chi$. Assuming $\chi$=1.5, the growth rate for the 1$^{st}$ head-tail mode is calculated to be $\mathrm{Im}(\Omega)/\omega_0$=0.01$Q_s$, which is 20-40 times smaller than the Landau damping rate for this mode. Thus, there is a significant safety factor for beam coherent stability.

For the extraction energy, the synchrotron frequency goes fast to its minimal value at the very end of the cycle. This results in $q$=10 at extraction, making the first 3-4 head-tail modes formally unstable. However, the growth times of these modes are calculated as 20 ms or larger and are too long to cause concern, since the entire acceleration time is 50 ms.

The longitudinal microwave stability threshold (Keil-Schnell) is calculated as 10-20 times above the nominal beam current (at all energies), so this instability should not be a problem.

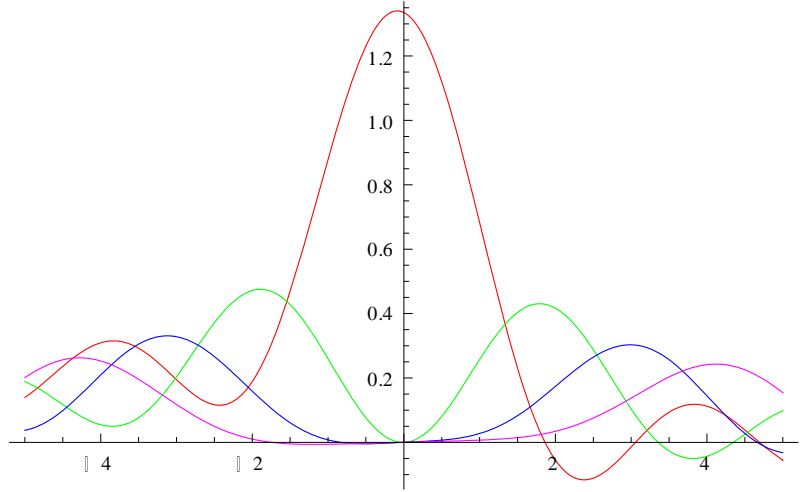

**Figure 5:** Mode factors $\Gamma$ as functions of the head-tail phase $\chi$ for the head-tail modes 0 (red), 1 (green), 2 (blue) and 3 (magenta), and the coupled-bunch mode number -19.

## V   RCS RF systems

Figure 6 presents time dependences of beam and RF system parameters during the acceleration and Table II shows the main parameters of the RF system. The maximum RF voltage for a cavity is chosen so that the nominal RF voltage could be achieved even if 2 of 16 cavities do not operate.

The beam-induced voltage significantly exceeds the RF system voltage required for the beam acceleration and capture. Even at the RF voltage maximum of 1.2 MV, the beam induced voltage (on resonance) exceeds the required RF voltage by 1.5 times. Cavity coupling is adjusted to minimize peak RF power during accelerator cycle. Together with optimal adjustment of cavity detuning in during accelerating cycle that enables a reduce RF power per cavity to 80 kW if only 14 cavities operate.

**Table II: Main Parameters of RCS RF systems**

| Harmonic number | 98 |
|---|---|
| Maximum voltage, MV | 1.5 |
| Minimum voltage, kV | 20 |
| Frequency sweep, MHz | 50.33-52.81 |
| Number of cavities | 16 |
| Loaded shunt impedance, k$\Omega$ | 50 |
| Peak RF power (per cavity, 14 cavity operation) | 80 |

The injection to the Recycler, and, subsequently, to MI requires injection gaps of 3 buckets in the bunch structure of MI. The beam abort and the extraction from the MI at the maximum energy require an abort gap of 45 buckets in MI. Therefore, the extraction gap in RCS is chosen to be 3 buckets, same as in the present Booster.

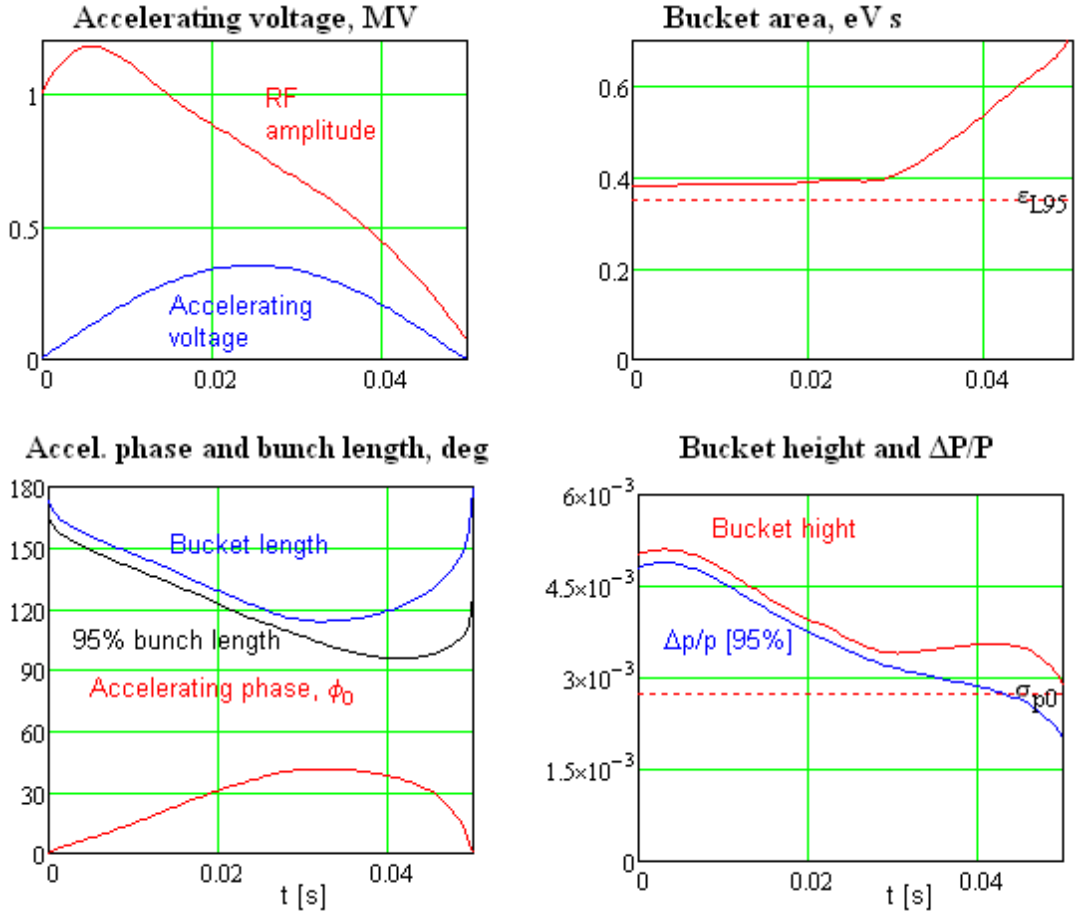

**Figure 6:** Beam and RF system parameters during acceleration. The longitudinal emittance is equal to 0.35 eV s and is not changed during the acceleration.

## VI  RCS Technical Systems

Other technical details for injection, extraction, collimation, magnet, and vacuum systems can be found in Refs. [1, 2].

## Summary

We have described a cost-effective 2-8 GeV Rapid Cycling Synchrotron, capable of delivering 340-kW beam power at 10 Hz and supporting the Fermilab Main Injector 2.4-MW operations. At 10 Hz, about one half of the RCS beam cycles will be used for the MI injection, while the other half can be used for other 8-GeV experiments, including an existing beam delivery to the Fermilab Muon campus via the Recycler. The cost-effectiveness is achieved by reducing the size of the vacuum chamber, which in turn, allows for more compact magnets and less complex rf cavities, and by reusing the Recycler ring. The cost estimates, performed earlier, show that the cost of RF system (includes RF cavities and RF power) makes the major contribution to the total macine cost.

## Acknowledgements

We would like to thank many contributors to this concept, including Steve Holmes, Paul Derwent, Alexey Burov, Nikolay Mokhov, Igor Rakhno, Vladimir Kashikhin, Brian Chase, and John Reid. Fermilab is Operated by Fermi Research Alliance, LLC under Contract No. DE-AC02-07CH11359 with the U.S. Department of Energy.